# Supplementary Material for the Information Sciences Paper: An Experimental Study of Hyper-Heuristic Selection and Acceptance Mechanism for Combinatorial *t*-way Test Suite Generation

**KAMAL Z. ZAMLI and FAKHRUD DIN**
*IBM Centre of Excellence*
*Faculty of Computer Systems and Software Engineering*
*Universiti Malaysia Pahang*
*Lebuhraya Tun Razak, 26300 Kuantan, Pahang Darul Makmur, Malaysia*
*Email: kamalz@ump.edu.my*

**GRAHAM KENDALL**
*School of Computer Science*
*University of Nottingham Malaysia Campus*
*Jalan Broga, 43500 Semenyih, Selangor Darul Ehsan, Malaysia*

**BESTOUN S. AHMED**
*Department of Computer Science*
*Faculty of Electrical Engineering*
*Czech Technical University Karlovo n'am. 13, 121 Pracha 2, Czech Republic*

## 1. Introduction

Software testing relates to the process of accessing the functionality of a program against some defined specifications. To ensure conformance, test engineers often generate a set of test cases to validate against the user requirements.

Owing to the growing complexity of software and its increasing diffusion into various application domains, it is no longer unusual for a software project to have testing teams in more than one location or even distributed over many continents. Owing to the intertwined dependencies of many software development activities and their geographical and temporal issues, there are potentially many overlapping test cases which can cause unwarranted redundancies across the shared modules (i.e. a test for one requirement may be covered by more than one test).

In this paper, we explore the application of our newly developed hyper-heuristic, called Fuzzy Inference Selection (FIS), for addressing test redundancy reduction problem highlighted in [1]. This paper presents the supplementary results for the paper – "An Experimental Study of Hyper-Heuristic Selection and Acceptance Mechanism for Combinatorial *t*-way Test Suite Generation" published in Elsevier's Information Sciences.

The paper is organized as follows. Section 2 presents test redundancy reduction problem. Section 3 describes the FIS implementation. Section 4 presents our benchmarking experiments. Finally, section 5 discusses our experimental observations.

## 2. The Test Redundancy Reduction Problem

The test redundancy problem can be expressed as follows:

> **Given:** A test suite $T_S$, a set of test requirements Req= {$req_1$, $req_2$, ... $req_n$} that must be covered to provide the desired test coverage of the program, and subset of $T_S$ = {$T_1$, $T_2$, ... $T_n$}, one associated with each of the $req_i$'s such that any one of the test cases $t_j$ belonging to $T_i$ can be used to test $req_i$.

> **Problem:** Find representative set of test cases, $T_M$, based on permutation of $T_S$ that satisfies all of the $req_i$'s.

## 3. FIS Fuzzy Rules for Test Redundancy Reduction Problem

FIS represents our implementation of a selection hyper-heuristic. FIS is derived from our earlier work on ISR described [2]. Specifically, like ISR, FIS adopts three operators (i.e. improvement, diversification intensification) based on a Hamming distance measure. Recall that the improvement operator checks for improvements in the quality of the objective function. The diversification operator measures how diverse the current and the previously generated solutions are against the population of potential candidate solutions. Finally, the intensification operator evaluates how close the current and the previously generated solutions are against the population of solutions. As far as FIS is concerned, these three operator measures are used as input variables. Unlike ISR which uses strict Boolean logic, the proposed FIS also accepts partial truth (i.e. based on some degree of membership) allowing more objective control to maintain or potentially change any particular search operator during runtime. In this case, the operator selection is set as the output variable.

Owing to its performance, we opt for Mamdani inference [3, 4] with trapezoidal membership functions and centroid defuzzification for our implementation. Figure 1 summarizes our implementation as a block diagram.

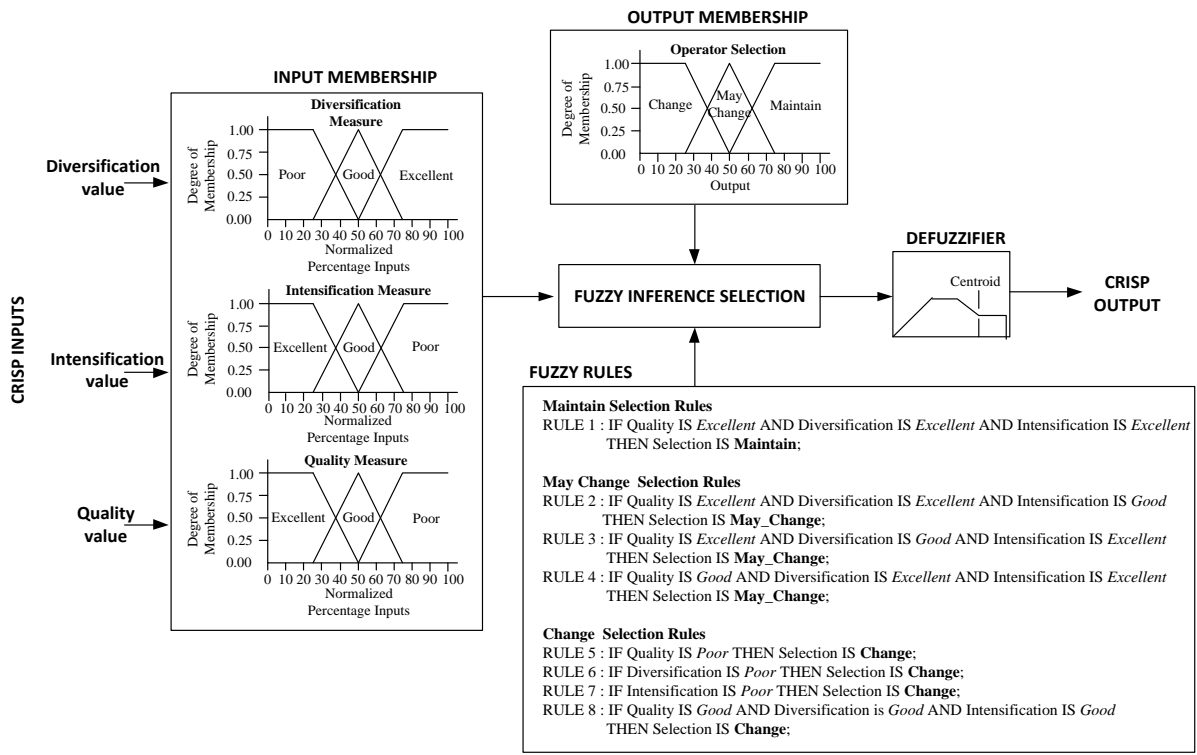

Figure 1. Fuzzy Inference Selection for Test Redundancy Reduction Problem

With the exception of the quality membership function which sometimes require reassignment between linguistic term *Excellent* and *Poor* depending on the (minimum or maximum) objective values, the rest of fuzzy rules are sufficiently general and potentially be applicable in other optimization problem.

## 4. The Benchmark Experiments

In order to demonstrate its performances, we have decided to benchmark FIS against existing work namely HGS[1], GE [5], GRE [6] and tReductSA [7]. Specifically, we have divided our experiments into two main parts. In the first part, we have adopted 3 main experiments which are reported in [6]. For the first experiment, we take the tests and requirement mapping involving Req={ $req_1$, $req_2$, ... $req_{19}$} and T={$t_1,t_2,t_3,t_4,t_5,t_6,t_7$}. In the second experiment, we adopt the tests and requirement mapping with Req={ $req_1$, $req_2$, ... $req_{19}$} and T={$t_1,t_2,t_3,t_4,t_8,t_9$}. Finally, in the third experiment, we use the tests and requirement mapping with Req={ $req_1$, $req_2$, ... $req_{19}$} and T={$t_1,t_3,t_4,t_5,t_6,t_8,t_{10},t_{11},t_{12}$ }. The detailed configurations for the first part are shown in Table 1.

**Table 1.** Benchmark Case Studies for Part 1

|  | Experiment 1 | Experiment 2 | Experiment 3 |
|---|---|---|---|
| $Req_i$ | $T_n$ | $T_n$ | $T_n$ |
| $req_1$ | $\{t_1,t_2,t_3,t_4,t_5,t_6,t_7\}$ | $\{t_1,t_2,t_3,t_4,t_8,t_9\}$ | $\{t_1,t_3,t_4,t_5,t_6,t_8,t_{10},t_{11},t_{12}\}$ |
| $req_2$ | $\{t_1,t_2,t_3,t_4,t_5,t_6,t_7\}$ | $\{t_1,t_2,t_3,t_4,t_8,t_9\}$ | $\{t_1,t_3,t_4,t_5,t_6,t_8,t_{10},t_{11},t_{12}\}$ |
| $req_3$ | $\{t_1,t_2,t_3,t_4,t_5,t_6,t_7\}$ | $\{t_1,t_2,t_3,t_4,t_8,t_9\}$ | $\{t_1,t_3,t_4,t_5,t_6,t_8,t_{10},t_{11},t_{12}\}$ |
| $req_4$ | $\{t_1,t_2,t_3,t_4,t_5,t_6,t_7\}$ | $\{t_1,t_2,t_3,t_4,t_8,t_9\}$ | $\{t_1,t_3,t_4,t_5,t_6,t_8,t_{10},t_{11},t_{12}\}$ |
| $req_5$ | $\{t_1,t_2,t_5,t_7\}$ | $\{t_1,t_2,t_9\}$ | $\{t_1,t_5,t_{10},t_{11},t_{12}\}$ |
| $req_6$ | $\{t_2,t_3,t_4,t_6\}$ | $\{t_2,t_3,t_4,t_8,t_9\}$ | $\{t_3,t_4,t_6,t_8,t_{10},t_{12}\}$ |
| $req_7$ | $\{t_1,t_7\}$ | $\{t_1\}$ | $\{t_1,t_{10},t_{12}\}$ |
| $req_8$ | $\{t_2,t_5\}$ | $\{t_2,t_9\}$ | $\{t_5,t_{11}\}$ |
| $req_9$ | $\{t_1,t_7\}$ | $\{t_1\}$ | $\{t_1,t_{10},t_{12}\}$ |
| $req_{10}$ | $\{t_1,t_2,t_5,t_7\}$ | $\{t_1,t_2,t_9\}$ | $\{t_1,t_5,t_{10},t_{11},t_{12}\}$ |
| $req_{11}$ | $\{t_2,t_3\}$ | $\{t_2,t_3,t_8\}$ | $\{t_3,t_8,t_{10}\}$ |
| $req_{12}$ | $\{t_3,t_4,t_6\}$ | $\{t_3,t_4,t_8,t_9\}$ | $\{t_3,t_4,t_6,t_8,t_{12}\}$ |
| $req_{13}$ | $\{t_2,t_3\}$ | $\{t_2,t_3,t_8\}$ | $\{t_3,t_8,t_{10}\}$ |
| $req_{14}$ | $\{t_2,t_3\}$ | $\{t_2,t_3,t_8\}$ | $\{t_3,t_8,t_{10}\}$ |
| $req_{15}$ | $\{t_3,t_4,t_7\}$ | $\{t_3,t_4,t_9\}$ | $\{t_3,t_4,t_{12}\}$ |
| $req_{16}$ | $\{t_4,t_6\}$ | $\{t_4,t_8\}$ | $\{t_4,t_6,t_8\}$ |
| $req_{17}$ | $\{t_3,t_4\}$ | $\{t_3,t_4,t_9\}$ | $\{t_3,t_4,t_{12}\}$ |
| $req_{18}$ | $\{t_3,t_4\}$ | $\{t_3,t_4,t_9\}$ | $\{t_3,t_4,t_{12}\}$ |
| $req_{19}$ | $\{t_4,t_6\}$ | $\{t_4,t_8\}$ | $\{t_4,t_6,t_8\}$ |

**Table 2.** Benchmark Case Studies for Part 2

|  | Experiment 4 | Experiment 5 |
|---|---|---|
| $Req_i$ | $T_n$ | $T_n$ |
| $req_1$ | $\{t_0,t_3,t_7,t_{18},t_{29}\}$ | $\{t_0,t_3,t_7,t_{18},t_{19},t_{29}\}$ |
| $req_2$ | $\{t_3,t_{16},t_{22}\}$ | $\{t_1,t_2,t_3,t_6,t_{12},t_{16},t_{22},t_{24}\}$ |
| $req_3$ | $\{t_0,t_2,t_{25},t_{27}\}$ | $\{t_0,t_2,t_{25},t_{27}\}$ |
| $req_4$ | $\{t_{11},t_{30}\}$ | $\{t_{11},t_{30}\}$ |
| $req_5$ | $\{t_1,t_4,t_8,t_{14},t_{25}\}$ | $\{t_1,t_4,t_8,t_{14},t_{25}\}$ |
| $req_6$ | $\{t_9,t_{14},t_{19},t_{24}\}$ | $\{t_9,t_{14},t_{19},t_{24}\}$ |
| $req_7$ | $\{t_5,t_{10},t_{21}\}$ | $\{t_5,t_{10},t_{21}\}$ |
| $req_8$ | $\{t_4,t_{20}\}$ | $\{t_4,t_{20}\}$ |
| $req_9$ | $\{t_7,t_{17},t_{24},t_{26}\}$ | $\{t_7,t_{17},t_{24}\}$ |
| $req_{10}$ | $\{t_6,t_{15},t_{29}\}$ | $\{t_{15},t_{29}\}$ |
| $req_{11}$ | $\{t_{10},t_{15},t_{23}\}$ | $\{t_{10},t_{15},t_{23}\}$ |
| $req_{12}$ | $\{t_1,t_6\}$ | $\{t_1,t_6\}$ |
| $req_{13}$ | $\{t_4\}$ | $\{t_6\}$ |
| $req_{14}$ | $\{t_2,t_8,t_{13},t_{16},t_{23}\}$ | $\{t_2,t_8,t_{13},t_{16},t_{23}\}$ |
| $req_{15}$ | $\{t_{28}\}$ | $\{t_{20},t_{28}\}$ |
| $req_{16}$ | $\{t_{22},t_{28}\}$ | $\{t_0,t_{18},t_{22}\}$ |
| $req_{17}$ | $\{t_{17},t_{29}\}$ | $\{t_{17},t_{29}\}$ |
| $req_{18}$ | $\{t_5,t_{20}\}$ | $\{t_5,t_{20}\}$ |
| $req_{19}$ | $\{t_9,t_{25}\}$ | $\{t_9,t_{25}\}$ |
| $req_{20}$ | $\{t_{12}\}$ | $\{t_{10},t_{12}\}$ |
| $req_{21}$ | $\{t_9,t_{28},t_{30}\}$ | $\{t_9,t_{28},t_{30}\}$ |
| $req_{22}$ | $\{t_3,t_{24}\}$ | $\{t_3,t_{24}\}$ |
| $req_{23}$ | $\{t_0,t_{30}\}$ | $\{t_0,t_5,t_{30}\}$ |
| $req_{24}$ | $\{t_5,t_8,t_{11},t_{26},t_{27}\}$ | $\{t_5,t_8,t_{11},t_{13},t_{26},t_{27}\}$ |

For the second part, we have adopted two new experiments (i.e. with slightly larger configuration than that of the first part) involving Req={ $req_1$, $req_2$, ... $req_{24}$} and T={$t_0,t_2,...,t_{30}$}. Unlike the first part (where we use the available result from [6]), no results are readily available for comparison in this case. Hence, for both experiments, we compare FIS against our own implementation of tReductSA and GE [5]. The detail configurations for the second part are shown in Table 2.

In our FIS experiments, we adopt the population size = 20 and the maximum number of iterations =*100*. FIS is implemented in the Java programming language. In the case of tReductSA, the simulated annealing parameters are $\alpha$ = 0.990, $T_{initial}$ = 2984.975, and $T_{final}$ =0.000. Similar to tReductSA, we run our FIS implementation for 15 times and recorded our best test sequence results (which could be in any order).

Our experimental platform comprises of a PC running Windows 10, CPU 2.9 GHz Intel Core i5, 16 GB 1867 MHz DDR3 RAM and a 512 MB of flash HDD. The complete benchmarking results for both parts of the comparison are tabulated in Table 3 and Table 4 respectively.

Table 3. Benchmarking Results for Part 1

|  | Experiment 1 | Experiment 2 | Experiment 3 |
|---|---|---|---|
| **GRE** | {$t_2,t_4,t_1(t_7)$} <br> Reduction = 57% | {$t_1,t_3,t_2(t_9)$, $t_4(t_8)$} <br> Reduction = 33% | {$t_5(t_{11}),t_3,t_{10}(t_{12}),t_4(t_8)$} <br> Reduction = 50% |
| **GE** | {$t_3,t_1(t_7),t_4(t_6),t_2(t_5)$} <br> Reduction = 43% | {$t_1,t_3,t_2(t_9),t_4(t_8)$} <br> Reduction = 33% | {$t_{12},t_8,t_5(t_{11})$} <br> Reduction = 66% |
| **HGS** | {$t_3,t_1(t_7),t_4(t_6),t_2(t_5)$} <br> Reduction = 43% | {$t_1,t_4,t_2$} or {$t_1,t_8,t_9$} <br> Reduction = 50% | {$t_5(t_{11}),t_3,t_1(t_{10},t_{12}),t_4(t_6,t_8)$} <br> Reduction = 50% |
| **tReductSA** | {$t_2,t_4,t_1(t_7)$} <br> Reduction = 57% | {$t_1,t_4,t_2$} or {$t_1,t_8,t_9$} <br> Reduction = 50% | {$t_5(t_{11}),t_8,t_{12}$} or {($t_5(t_{11}),t_{10},t_4$} <br> Reduction = 66% |
| **FIS** | {$t_2,t_4,t_1(t_7)$} <br> Reduction = 57% | {$t_1,t_4,t_2$} or {$t_1,t_8,t_9$} <br> Reduction = 50% | {$t_5(t_{11}),t_8,t_{12}$} or {($t_5(t_{11}),t_{10},t_4$} <br> Reduction = 66% |

Table 4. Benchmarking Results for Part 2

|  | Experiment 4 | Experiment 5 |
|---|---|---|
| **GE** | *{$t_4,t_{28},t_{12},t_5,t_3,t_2,t_6,t_9,t_{17},t_{10},t_{11}$}* <br> *Reduction = 65%* | *{$t_6,t_0,t_5,t_9,t_4,t_{10},t_{17},t_2,t_3,t_{11},t_{15},t_{20}$}* <br> *Reduction = 61%* |
| **tReductSA** | *{$t_7,t_{17},t_{12},t_3,t_{25},t_6,t_{30},t_{28},t_{15},t_4,t_5,t_{24},t_{23}$}* <br> *Reduction = 58%* | *{$t_7,t_{29},t_{11},t_3,t_{16},t_{20},t_0,t_{10},t_9,t_6,t_8$}* or <br> *{$t_{29},t_{27},t_{18},t_{28},t_{20},t_{30},t_{10},t_9,t_6,t_8,t_{24}$}* <br> *Reduction = 64%* |
| **FIS** | *{$t_{23},t_0,t_{24},t_{16},t_{29},t_4,t_{12},t_9,t_5,t_1,t_{28}$}* or <br> *{$t_1,t_{16},t_{24},t_{12},t_4,t_{23},t_{25},t_0,t_{28},t_{29},t_5$}* or <br> *{$t_3,t_4,t_{28},t_6,t_{17},t_5,t_{23},t_{25},t_{12},t_{19},t_{30}$}* or <br> *{$t_5,t_{28},t_4,t_{29},t_{16},t_{30},t_{10},t_{24},t_{12},t_{25},t_6$}* or <br> *{$t_5,t_{24},t_{25},t_{30},t_6,t_{28},t_{12},t_{23},t_4,t_{16},t_{29}$}* <br> *Reduction = 65%* | *{$t_0,t_8,t_{24},t_{10},t_{20},t_6,t_{25},t_{30},t_{29}$}* or <br> *{$t_{30},t_8,t_6,t_{24},t_9,t_{20},t_0,t_{29},t_{10}$}* or <br> *{$t_6,t_{30},t_8,t_{22},t_{29},t_{20},t_{24},t_{10},t_{25}$}* <br> *Reduction = 71%* |

## 5. Experimental Observation

Reflecting on the work undertaken, a number of observations can be elaborated based on the obtained results.

Referring to Table 3, FIS, tReductSA and GRE outperforms both GE and HGS with 57% reduction in the first experiment generating the same set of test sequence (i.e. can be in any order). Although not optimal, the same set of test suite is also generated by GE and HGS. For the second experiment, FIS, tReductSA and HGS outperforms GRE and GE with 50% reduction. The same set of test sequence is generated by FIS, tReductSA and HGS. Similar with earlier case, although not the best, the same test sequence is also generated by GRE and GE. Finally, in the third experiment, FIS, tReductSA and GE also gives the same percentage of reduction with 66%.

Concerning the result for part 2 in Table 8, FIS and GE outperform tReductSA for with 65% and 58% reduction respectively for experiment 4. The test sequences for FIS are also more diversified than others. Concerning experiment 5, FIS gives the overall best with 71% reduction. tTReductSA comes as the runner up with 64% followed by GE with 61% .

Summing up, FIS has been able to match with the best results for all five experiments. In fact, in experiment 5, FIS gives the overall best as compared to tReductSA and GE.

**Acknowledgement**

The work reported in this paper is funded by the Science Fund Grant for the project titled: Constraints T-Way Testing Strategy with Modified Condition/Decision Coverage from the Ministry of Science, Technology, and Innovation (MOSTI), Malaysia.**References**